\documentclass[aps,prb,showpacs,showkeys]{revtex4}

\usepackage{graphicx}
\usepackage{dcolumn}
\usepackage{bm}
\usepackage{amsmath,amssymb,amsfonts}

\newcommand{\TG}{\mathcal{G}}
\newcommand{\TT}{\mathcal{T}_2}
\newcommand{\TF}{\mathit{\Phi}}
\renewcommand{\Im}{\mathfrak{Im}}
\renewcommand{\Re}{\mathfrak{Re}}

\begin{document}


\title{Exact diagrammatic approach for dimer-dimer scattering and bound states of three and four resonantly interacting particles.}


\author{I.V.~Brodsky}
    \affiliation{P.L. Kapitza Institute
    for Physical Problems, Kosygin street 2, Moscow, Russia, 119334}
\author{A.V.~Klaptsov}
    \affiliation{Russian Research Centre "Kurchatov Institute",
    Kurchatov square 1, Moscow, Russia, 123182}
\author{M.Yu.~Kagan}
    \email[Corresponding author: ]{kagan@kapitza.ras.ru}
    \affiliation{P.L. Kapitza Institute
    for Physical Problems, Kosygin street 2, Moscow, Russia, 119334}
\author{R.~Combescot}
    \affiliation{Laboratoire de Physique Statistique, Ecole Normale Sup\'{e}rieure, 24 rue Lhomond, 75231 Paris Cedex 05,
    France}
\author{X.~Leyronas}
    \affiliation{Laboratoire de Physique Statistique, Ecole Normale Sup\'{e}rieure, 24 rue Lhomond, 75231 Paris Cedex 05,
    France}



\date{\today}

\begin{abstract}
We present an exact diagrammatic approach for the dimer-dimer scattering problem
in two or three spatial dimensions, within the resonance approximation where
these dimers are in a weakly bound resonant state. This approach is first applied
to the calculation of the dimer-dimer scattering length $a_B$ in three spatial dimensions, 
for dimers made of two fermions in a spin-singlet state, with corresponding scattering 
length $a_F$ and the already known result $a_B=0.60\,a_F$ is recovered exactly. 
Then we make use of our approach to obtain new results in two spatial dimensions 
for fermions as well as for bosons. Specifically, we calculate bound state energies for
three $bbb$ and four $bbbb$ resonantly interacting bosons in two dimensions.
We consider also the case of resonant interaction between fermions and bosons
and we obtain the exact bound state energies of two bosons plus one fermion $bbf$, 
two bosons plus two fermions $bf_{\uparrow}bf_{\downarrow}$, and three bosons 
plus one fermion $bbbf$.

\end{abstract}


\maketitle

\section{Introduction}

Following the experimental realization of the Bose-Einstein
condensation in ultracold bosonic gases, together with its
intensive study, the physics of ultracold Fermi gases has taken off
recently with a strong development of experimental and theoretical
investigations within the last few years \cite{Levico:2004}. In particular, much advantage
has been taken of various Feshbach resonances which offer the
possibility observing experimentally the so called BEC-BCS crossover.
This has been done in particular in $^6$Li and in $^{40}$K.
In the weak coupling limit of small negative scattering length, which
is realized far away on one side of the resonance, the corresponding
weak attractive interaction between fermions leads to a BCS type
condensate of Cooper pairs. On the other side of the resonance, where
the scattering length is positive, weakly bound dimers, or molecules,
consisting of two different fermions are formed. When one goes far
enough of the resonance on this positive side, one obtains
a weakly interacting gas of these dimers, which may in particular
form a Bose-Einstein condensate, as it has been recently observed experimentally \cite{Greiner:2003,
Jochim:2003,Zwierlein:2003,Bourdel:2004}. 

In the present paper, motivated by the problem raised by the physics of this
dilute gas of composite bosons, we will deal with the dimer-dimer elastic
scattering and present an exact diagrammatic approach to its solution. 
This will be done by staying in the so-called resonance approximation
which is quite suited to the physical situation found with a Fesbach
resonance. It this case the (positive) scattering length greatly
exceeds the characteristic radius $r_0$ for the attractive interaction 
between fermionic atoms. A problem of this kind was first investigated
by Skorniakov and Ter-Martirosian
\cite{Skorniakov/Ter-Martirosian:1956} in the case of the 3-body
fermionic problem. They showed that the scattering length of a fermion on a weakly
bound dimer is determined by a single parameter, namely the two-body
scattering length $a_F$ between fermions, and it is equal to $1.18a_F$ in the
zero-range limit for the interatomic potential. A similar situation is found
in the case of four fermions, where the dimer-dimer scattering length is fully 
determined by this same scattering length $a_F$.

In a study of the crossover problem Haussmann
\cite{Haussmann:1993} calculated this scattering length
of composite bosons $a_B$ at the level of the Born approximation
and found it equal to $2a_F$. This result was later on much improved
by Pieri and Strinati \cite{Pieri/Strinati:2000}, who took into
account the repeated scattering of these composite bosons
in the ladder approximation. This diagrammatic
approach led them to a scattering length 
approximately equal to $a_B \simeq 0.75a_F$. However, this ladder
approximation is not exact, because it misses
an infinite number of other diagrams which in principle lead to
a contribution of the
same order of magnitude as those taken into account. Very recently
this problem has been solved exactly by
Petrov, Salomon, and Shlyapnikov \cite{Petrov/Salamon:2004,Petrov/Salamon:2005} 
who found for the scattering length of these composite bosons 
$a_B=0.6a_F$. This has been achieved by solving directly 
the Schr\"{o}dinger equation for four fermions, using the
well-known method of pseudopotentials. Here we will give an exact
solution of this scattering problem of two weakly bound dimers,
using a diagrammatic approach in the resonance approximation,
which can be seen as a bridge between the approach of Pieri and
Strinati \cite{Pieri/Strinati:2000} and the exact result of
Petrov, Salomon, and Shlyapnikov \cite{Petrov/Salamon:2004,Petrov/Salamon:2005}.

In order to show the strength and the versatility of our approach,
we make use of it to obtain new results for various
systems, in the two-dimensional (2D) case which is of interest not
only for cold gases, but also for high $T_c$ superconductivity.
Specifically we consider first a system
of resonantly interacting bosons. We calculate exactly the three
bosons $bbb$ and four bosons $bbbb$ bound state energies in this
case. We also make use of our approach for the study of 2D bosons
interacting resonantly with fermions. In this case we calculate exactly
the bound state energies of two bosons
plus one fermion $bbf$, two bosons plus two fermions
$bf_{\uparrow}bf_{\downarrow}$, and three bosons plus one fermion
$bbbf$. In this respect the present paper is in the line of previous results
obtained by of some of us. Indeed
the possibility of two fermions \cite{Kagan/Rice:1994, Kagan/Fresard:1998}
$ff$ and two bosons \cite{Kagan/Efremov:2002}
$bb$ pairing  was predicted, as well as the creation \cite{Kagan/Brodsky:2004c}
of a composite fermion $bf$ in resonantly
interacting $(a\gg r_0)$ 2D Fermi-Bose mixtures.

\section{Three particles scattering}

\begin{figure*}
    \includegraphics[width=15pc]{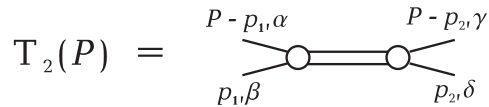}
    \caption{\label{fig:Gamma2} The graphic representation of the two
    particles
    vertex $T_2(P)$ (the four external propagators do not belong to $T_2(P)$).}
\end{figure*}

As a preliminary exercise we will rederive the result of
Skorniakov and Ter-Martirosian for the dimer-fermion scattering length $a_3$
using the diagrammatic method ~\cite{Bedaque/vanKolck:1998}.
Following Skorniakov and Ter-Martirosian, in the presence of the
weakly bound resonance level $-E_b$ (with $E_b>0$), we can limit ourselves to 
the zero-range interaction potential between fermions
in the scattering of these two particles . The two-fermion vertex  can be
approximated by a simple one-pole structure, which reflects the
presence of the s-wave resonance level in the spin-singlet state, and
is essentially given by the scattering amplitude, namely:
\begin{eqnarray}\label{2Vertex}
    T_{2\,\alpha\beta;\,\gamma\delta}(P) =T_2(P)\times
    (\delta_{\alpha,\,\gamma}\delta_{\beta,\,\delta} -
    \delta_{\alpha,\,\delta}\delta_{\beta, \, \gamma}) . (\delta_{\alpha,
    \uparrow}\delta_{\beta,\downarrow}+\delta_{\alpha,
    \downarrow}\delta_{\beta, \uparrow})=
    T_2(P)\chi(\alpha, \beta)\chi(\gamma, \delta),\\
    T_2(P) = \frac{4\pi}{m^{3/2}}\,
    \frac{\sqrt{E_b}+\sqrt{\mathbf{P}^2/4m-E}}{E
    - \mathbf{P}^2/4m+E_b} , \qquad
     \chi(\alpha,\beta)=\delta_{\alpha,
    \uparrow}\delta_{\beta,\downarrow}-\delta_{\alpha,
    \downarrow}\delta_{\beta, \uparrow},
\end{eqnarray}
where $P = \{\textbf{P},E\}$,  $E$ is the total frequency and
$\textbf{P}$ is the total momentum of incoming particles, $m$ is the
fermionic mass, $E_b = 1/ma_F^2$. Indices $\alpha, \beta$ and
$\gamma, \delta$ denote the spin states of incoming and outgoing
particles. The function $\chi(\alpha,\beta)$ stands for the spin
singlet state. We will draw this vertex in the way, shown on
Fig. \ref{fig:Gamma2}, where the double line can be regarded as a
propagating dimer.

\begin{figure*}
    \includegraphics[width=20pc]{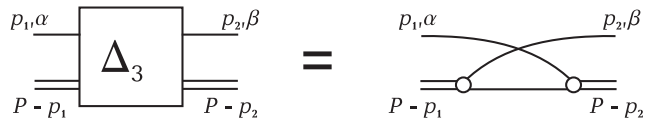}
    \caption{\label{fig:Gamma3_0} The graphic representation of the simplest
    dimer-fermion scattering process $\Delta_3$  (the two external fermion 
    propagators and the two external dimer propagators do not belong to $\Delta_3$).}
\end{figure*}

\begin{figure*}
    \includegraphics[width=30pc]{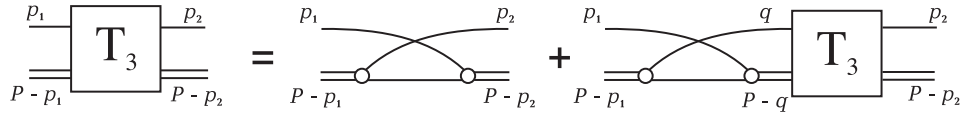}
    \caption{\label{fig:Gamma3} The diagrammatic representation of the
    equation for the full dimer-fermion scattering vertex $T_3$.}
\end{figure*}

The simplest process that contributes to dimer-fermion interaction
is the exchange of a fermion. We denote the corresponding vertex
as $\Delta_3$ and it is
described by the diagram on Fig. \ref{fig:Gamma3_0}. Its
analytical expression reads
\begin{equation}
    \label{3VertexZero} \Delta_{3\,\alpha, \beta}(p_1,p_2;P) =
    -\delta_{\alpha,\beta}\,G(P-p_1-p_2),
\end{equation}
where $G(p) = 1/\left(\omega-\mathbf{p}^2/2m+i0_{+}\right)$ is the bare
fermion Green's function. The minus sign in the right hand side of
Eq.\eqref{3VertexZero} comes from the permutation of the two fermions. In
order to obtain the full dimer-fermion scattering vertex $T_3$ we
need to sum up all possible diagrams with indefinite number of $\Delta_3$ blocks.
In the present case these diagrams have a ladder structure. It is obvious
that the spin projection is conserved in every order in $\Delta _3$
and thus $T_{3\,\alpha,\beta}=\delta_{\alpha,\beta}\,T_3$. An
equation for $T_3$ will have the  diagrammatic representation
shown in Fig.~\ref{fig:Gamma3}. It is obtained by writing that either the simplest
exchange process occurs alone, or it is followed by any other process.
In analytical form it reads
\begin{equation}\label{3Vertex}
    T_3(p_1,p_2;P)=-G(P-p_1-p_2)- \sum_{q}G(P-p_1-q)
    G(q)\, T_2(P -q)\; T_3(q, p_2; P),
\end{equation}
where $\sum\limits_q\equiv i \int d^3qd\Omega/(2\pi)^4$.
We can integrate out the frequency $\Omega$ in Eq.\eqref{3Vertex}
by closing the integration contour in the lower half-plane, since
both $T_2(P -q)$ and $T_3(q, p_2; P)$ are analytical functions of $\Omega$ in
this region (this property for $T_3(p_1, p_2; P)$ results from Eq.\eqref{3Vertex} itself).
Hence only the "on the shell" value $T_3(\{{\bf q},q^{2}/2m\}, p_2; P)$ comes in the right-
hand side of Eq.\eqref{3Vertex}.
Moreover, if we are interested in the low-energy
s-wave dimer-fermion scattering length $a_3$, we have to put $P
= \{\mathbf{P},E\} = \{\mathbf{0},-E_b\}$ and $p_2=0$. Hence Eq.\eqref{3Vertex}
reduces to an equation for the "on the shell" value of $T_3(p_1, p_2; P)$.
Taking into account
the standard relation between $T$-matrix and scattering amplitude (with
reduced mass) and the fact that, from Eq.\eqref{2Vertex}, $T_2$ has
an additional factor $8\pi /(m^{2}a_{F})$ compared to a standard boson
propagator, we find that the full vertex
$T_3$ is connected with $a_3$ by the following relation:
\begin{equation}
    \left(\frac{8\pi}{m^2a_F}\right)T_3\left(0, 0; \{{\bf 0},-E_b\}\right)=
    \frac{3 \pi}{m}\,a_3.
\end{equation}
This leads to introduce a new function $a_3(\mathbf{k})$ defined by
\begin{equation}
    a_3(\mathbf{k})=\frac{4}{3m}\left(\sqrt{mE_b}+\sqrt{3k^2/4+mE_b}\right)\;T_3\left(\{\mathbf{k},k^2/2m\},0;\{{\bf 0}, -E_b\}\right).
\end{equation}
and substituting it in Eq.\eqref{3Vertex}, we obtain
Skorniakov - Ter-Martirosian equation for the scattering
amplitude:
\begin{equation}\label{a_3}
    \frac{(3/4)\,a_3({\mathbf k})}{\sqrt{mE_b}+\sqrt{3k^2/4+mE_b}}
    =\frac{1}{k^2+mE_b}-4\pi\int\frac{d{\bf q}}{(2\pi)^3}\,\frac{a_3(\mathbf{q})}
    {q^2\,(k^2+q^2+\mathbf{k.q} +mE_b)}.
\end{equation}
Solving this equation one obtains the well known result
\cite{Skorniakov/Ter-Martirosian:1956} for
the dimer-fermion scattering length $a_3 = a_3(0) = 1.18 a_F$.

\section{Dimer - dimer scattering}
By now we can proceed to the problem of the dimer-dimer scattering.
This problem was previously solved by Petrov et
al.~\cite{Petrov/Salamon:2004,Petrov/Salamon:2005} via studying Schr\"{o}dinger
equation for  a 4-fermions wave function. Our diagrammatic
approach is conceptually close to Petrov's one.  Its basic point is that
it requires the
introduction of a special vertex which describes an interaction of
one dimer as a single object with the two fermions constituting
the other dimer.

\begin{figure*}
    \includegraphics[width=40pc]{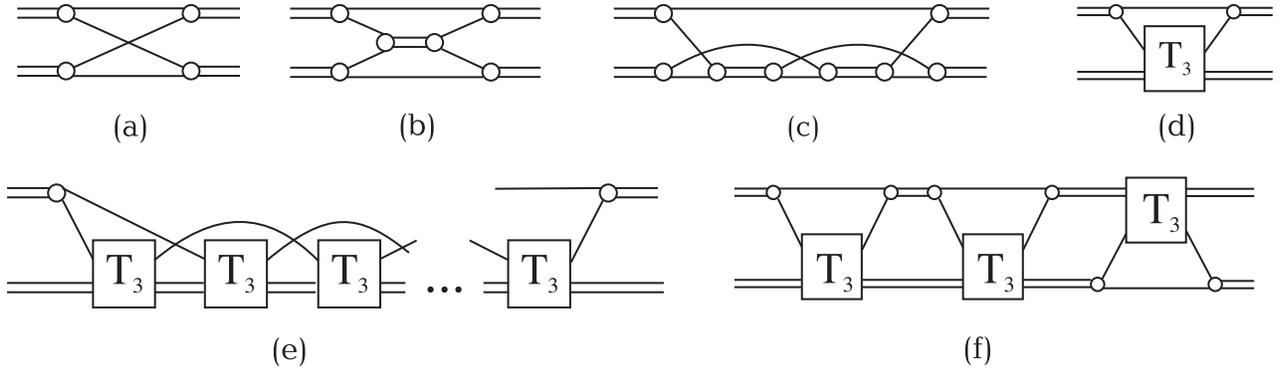}
    \caption{\label{fig:Gamma4_0} The graphic representation of the
    dimer-dimer scattering processes contributing to $T_4$.}
\end{figure*}

Let us investigate all the possible types of diagrams that contribute
to the dimer-dimer scattering vertex $T_4$. In this process both
dimers are temporarily "broken" in their fermionic components,
which means that the fermions of one dimer exchange and/or interact
with the fermions of the other dimer. The simplest process
is an exchange of fermions by two dimers  shown on
Fig.~\ref{fig:Gamma4_0}{\it a}. More complicated diagrams are composed
by introducing intermediate interactions between exchanging
fermions (see Fig.~\ref{fig:Gamma4_0}{\it b,c}). As long as one of the fermions
does not interact or exchange with the other ones, all these complications
can be summed up in the $T_3$ block (see
Fig.~\ref{fig:Gamma4_0}{\it d}) which describes, as we have seen
in the preceding section, the scattering of a
fermion on a dimer. Furthermore we may exchange bachelor fermions
participating in the $T_3$ scattering. The resulting series has
the diagrammatic structure shown on Fig.~\ref{fig:Gamma4_0}{\it e}.
This series describes a "bare" interaction between dimers. The
last obvious step is to compose ladder type diagrams from this
"bare" interaction. A typical ladder diagram is shown on
Fig.~\ref{fig:Gamma4_0}{\it f}. These general ladder diagrams
describe all possible processes which contribute to the dimer-dimer scattering.

The fact that the $T_4$ vertex should be expressed in terms of $T_3$
was first noticed by Weinberg in his work on multiparticle
scattering problems~\cite{Weinberg:1964}. Note that a calculation
of the diagrams shown on Fig.~\ref{fig:Gamma4_0}{\it e, f} requires
information about an off-shell matrix $T_3$, that is about a matrix
with arbitrary relation between frequencies and momenta of
incoming and outgoing particles.  On the other hand, for the
calculation of the dimer-fermion scattering length $a_3$ in
Eq.\eqref{a_3}, only the simpler on-shell structure of $T_3$ is
required as we have seen in the preceding section. Luckily, as we 
will see now, we can exclude $T_3$  from our considerations and
express $T_4$  only in terms of $T_2$. By doing this we reduce the
number of integral equations required for the calculation of the
dimer-dimer scattering length $a_4$.

\begin{figure*}
    \includegraphics[width=20pc]{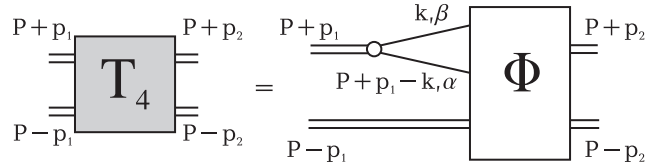}
    \caption{\label{fig:PhitoT_4}Diagrammatic representation of the relation
    between the full dimer-dimer scattering matrix $T_4$ and the vertex $\Phi$.}
\end{figure*}

Since, as we have just seen, it is impossible to construct a closed 
equation for the dimer-dimer scattering vertex $T_4$, we wish to find an
alternative way for taking into account in one equation  all
the diagrams contributing to dimer-dimer scattering. Inspired by
the work of Petrov et al.~\cite{Petrov/Salamon:2004,Petrov/Salamon:2005} 
and looking at the diagrams we have considered above, we are naturally lead to
look for a special vertex that describes the interaction of two fermions, 
constituting the first dimer, with the second dimer taken as a single
object. This vertex would be the sum of all
diagrams with two fermions  and one dimer as incoming lines. It
would be natural to suppose that these diagrams should have the
same set of outgoing -- two fermionic and one dimer -- lines.
However in this case there will be a whole set of disconnected
diagrams contributing to our sum that describe interaction of a
dimer with only one fermion. As it was pointed out by
Weinberg~\cite{Weinberg:1964}, one can construct a good integral
equation of Lippmann-Schwinger type only for connected class of
diagrams. Thus we are forced to pay our attention to the
vertex $\Phi_{\alpha\beta}(q_1,q_2;p_2, P)$ corresponding to the
sum of all diagrams with one incoming dimer, two incoming
fermionic lines and two outgoing dimer lines (see
Fig.~\ref{fig:PhitoT_4}). This is also quite natural from our view point
since, in our scattering problem we are interested in a 
final state with two outcoming dimers. Indeed
once this vertex $\Phi_{\alpha\beta}(q_1,q_2;p_2, P)$ is known, it is
straightforward to calculate the
dimer-dimer scattering vertex $T_4(p_1,p_2;P)$ which is given by:
\begin{equation}\label{PhitoT_4}
    T_4(p_1,p_2;P)=\frac{1}{2}\sum_{k;\,\alpha,\beta}\chi(\alpha,\beta)
    G(P+p_1-k)G(k)\Phi_{\alpha\beta}(P+p_1-k,k;p_2, P).
\end{equation}
The corresponding diagrammatic representation is given in 
Fig.~\ref{fig:PhitoT_4}. One can readily verify that, in any order
of interaction, $\Phi$ contains only connected diagrams.

\begin{figure*}
    \includegraphics[width=30pc]{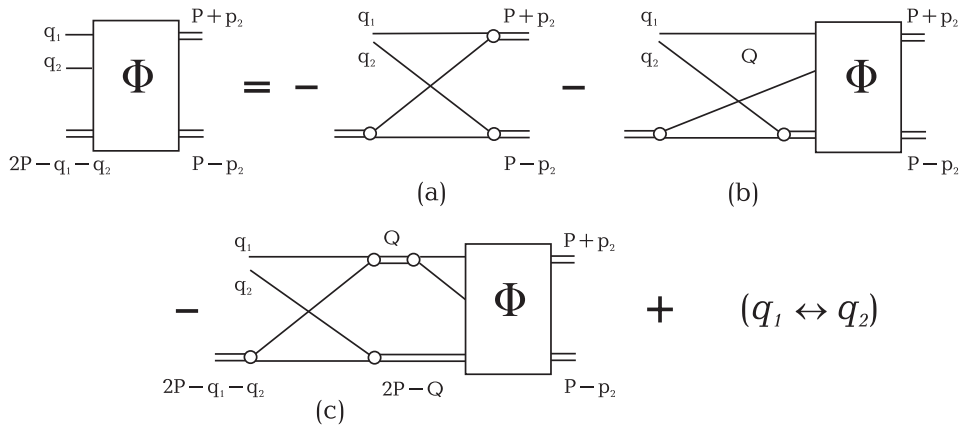}
    \caption{\label{fig:Phi} The diagrammatic representation of the integral equation
    for the function $\Phi$ introduced by dimer-dimer scattering .}
\end{figure*}

The spin part of the vertex $\Phi_{\alpha, \beta}$ has the simple form
$\Phi_{\alpha, \beta}(q_1,q_2;P,p_2) = \chi(\alpha,
\beta)\Phi(q_1,q_2;P,p_2)$. The diagrammatic representation of the
equation for $\Phi$ is given in Fig.~\ref{fig:Phi}.  One can
assign some "physical meaning" to the processes described by these
diagrams. The diagram of Fig.~\ref{fig:Phi}{\it a} represents the
simplest exchange process in a dimer-dimer interaction. The
diagram of Fig.~\ref{fig:Phi}{\it b} accounts for a more complicated
nature of a "bare" dimer-dimer
interaction. Finally the diagram of Fig.~\ref{fig:Phi}{\it c} allows for
a multiple dimer-dimer scattering via a "bare" interaction (it
generates ladder-type diagrams analogous to those of
Fig.~\ref{fig:Gamma4_0}{\it f}). The last term in Fig.~\ref{fig:Phi}
means that we should add another set of three diagrams analogous to
those of Fig.~\ref{fig:Phi}{\it a,~b,~c} but with the two incoming fermions ($q_1$
and $q_2$) exchanged. The diagrammatic representation translates into the
following analytical equation for the vertex $\Phi$:
\begin{multline}\label{Phi}
    \Phi(q_1,q_2;p_2, P)=-G(P-q_1+p_2)G(P-q_2-p_2)-
    \sum_{k}G(k)G(2P-q_1-q_2-k)T_2(2P-q_1-k)\Phi(q_1,k;p_2, P)    \\
    -\frac{1}{2}\sum_{Q, k}G(Q-q_1)G(2P-Q-q_2) T_2(2P-Q)T_2(Q)G(k)G(Q-k)\Phi(k,Q-k;p_2, P)
    +(q_1\leftrightarrow q_2).
\end{multline}
Finally let us also indicate that it is possible to rederive the same set of equations,
purely algebraically, by taking a complementary point of view. Instead of focusing, 
as we have done, on the free fermions lines as soon as a dimer is "broken", we can 
rather keep track of the fermions which make up a dimer. This leads again automatically 
to introduce the vertex $\Phi(q_1,q_2;p_2, P)$. Then Eq.\eqref{Phi} is recovered when
one keeps in mind that, after breaking dimers, one may have propagation of a
single dimer and two free fermions before another break (this corresponds to the second
term in the right hand side of Eq.\eqref{Phi}). Alternatively one may also have the 
propagation of two dimers, which leads to the third term in Eq.\eqref{Phi}.

Coming back now more specifically to our problem, we can put
$p_2 = 0$ and $P = \{\mathbf{0}, -E_b\}$ since we are looking for 
an s-wave scattering length. At this point we have a single
closed equation for the vertex $\Phi$ in momentum representation,
which we believe is analogous to Petrov {\it et al} equation in
coordinate representation. To make this analogy more prominent we
have to exclude frequencies from the equation by integrating them out. 
However this exclusion requires some more technical mathematics and 
we leave it out for Appendix A.

The dimer-dimer scattering length is directly related to the full
symmetrized vertex $T_4(p_1,p_2;P)$. Just as in the preceding section,
taking also statistics into account, we have:
\begin{equation}
    \left(\frac{8\pi}{m^2a_F}\right)^2T_4(0,0;\{{\bf 0}-E_b\},0)=\frac{2\pi(2a_B)}{m}.
\end{equation}
If one skips the second term in Eq.~\eqref{Phi}, i.e. one omits
diagram Fig.\ref{fig:Phi}{\it b}, one will arrive at the ladder
approximation of Pieri and Strinati \cite{Pieri/Strinati:2000}. The exact
equation~\eqref{Phi} corresponds to the summation of all diagrams.
We have calculated the scattering length in the ladder
approximation and the scattering length derived from the exact
equation and obtained $0.78a_F$ and $0.60a_F$ respectively. 
Some details on our actual procedure are given in the next section. Thus
our results in the ladder approximation are in agreement with the
results \cite{Pieri/Strinati:2000} of Pieri and Strinati and, in
the general form, with the results of Petrov \textit{et. al}
\cite{Petrov/Salamon:2004,Petrov/Salamon:2005}. 
Note also that our approach allows one
to find the dimer-dimer scattering length in the 2D case (this
problem was previously solved by Petrov \textit{et. al}
\cite{Petrov/Baranov:2003}).

Finally we would like to mention that our results allow one to
find a fermionic Green's function, chemical potential and sound
velocity as a function of $a_F$ in the case of dilute superfluid
bose gas of dimers at low temperatures. The problem of dilute
superfluid bose gas of di-fermionic molecules was solved by
Popov~\cite{Popov:1966}, and later deeply investigated by Keldysh
and Kozlov~\cite{Keldysh/Kozlov:1968}. Those authors managed to
reduce the gas problem to a dimer-dimer scattering problem in
vacuum, but were unable to express the dimer-dimer scattering
amplitude in a single two-fermion parameter. A direct combination
of our results with those ones of Popov, Keldysh and Kozlov allows
one to get all the thermodynamical values of a dilute superfluid
resonance gas of composite bosons. Another interesting
subject for the application of our results will be a
high-temperature expansion for the thermodynamical potential and
sound velocity in the temperature region $T\sim T_{*}\sim E_b$,
where the composite bosons begin to appear.

\section{Practical implementation}\label{pract}

Let us give now some details on the way in which we have solved 
effectively the above equations. Actually we have dealed with two
problems, the scattering length calculation discussed above and
the bound states problem to be discussed below. Our two problems 
are quite closely related since, for the scattering length problem, we look 
for the scattering amplitude at zero outgoing wavevectors and energy for two dimers,
while for the bound states we look for divergences of this same scattering amplitude at negative
energy. As already indicated, in both cases the situation is somewhat
simplified with respect to the variables we have to consider, due to the specific problem we
handle.  First with respect to $P = \{{\bf P},E\}$, we have ${\bf P}={\bf 0}$ since we work naturally
in the rest frame of the four particles. Moreover, with respect to the total energy, $E= -\epsilon E_b$
is negative. Specifically $\epsilon =1 $ when we look for the scattering length. Or when we consider
bound states $\epsilon $ gives the energy of the bound states we are looking for. Next, with respect
to parameter $p_2 \equiv \{{\bf p}_2,{\bar p}_2\}$ which characterizes the outgoing dimers,  we will have naturally ${\bf p}_2={\bf 0}$ as we have said since we consider zero outgoing wavevectors. Since we will evaluate ${\bar p}_2$ on the shell, we have merely ${\bar p}_2=0$, and this parameter drops out. Hence in the following we do not write anymore explicitely the value of parameter $P$.
Both for the scattering length problem and the bound
states problem, we have followed two main routes.

In our first route, we have written a specific integral equation for $T_4(p_1,p_2)$, 
which is then solved numerically.
The details of our derivation for this integral equation are given in Appendix B. The kernel for this equation is itself obtained from a vertex $\Gamma$. The defining integral equation Eq.(\ref{Gamma}) for this vertex has been inverted numerically, by calculating the inverse matrix, to obtain the vertex $\Gamma(q_1,q_2;p_2)$. We 
have used \cite{NumericalRecipes} LU factorization and Gauss quadrature. The result has then been substituted in Eq.(\ref{Delta}) which gives the kernel $\Delta_4(p_1,p_2)$ coming in the integral equation Eq.(\ref{T_4}). The solution of this last equation is naturally also handled numerically, for 
example by finding the eigenvalues of the kernel for the bound states problem.

In our second route we have kept both functions $T_4$ and $\Phi$. In the following we do not write anymore the parameter $p_2$ which takes always the trivial value $p_2=0$, as explained above. Hence we are left with $T_4(p_1)$
which, because of rotational invariance, depends only on the energy 
${\bar p}_1$ and the modulus $|{\bf p}_1|$ of the momentum.  For brevity we denote this quantity 
$t_4(|{\bf p}_1|,{\bar p}_1)$. On the other hand it is shown in Appendix A that, in order to evaluate the
second term in the right-hand side of Eq.(\ref{Phi}), we need only the 
evaluation of $\Phi(q_1,q_2)$ on the shell, which we denote as $\phi({\bf q}_1,{\bf q}_2)$. 
It depends only on the three variables $|{\bf q}_1|$, $|{\bf q}_2|$ and the angle between these two vectors. Hence it
is enough to write Eq.(\ref{Phi}) only for $q_1$ and $q_2$ taking on the shell values. From Eq.(\ref{Phi})
this leads for $\phi({\bf q}_1,{\bf q}_2)$ to the following more convenient equation:
\begin{eqnarray}
\phi({\bf q}_1,{\bf q}_2) = - \; \frac{1}{(|E|+{\bf q}_1^{2}/m)(|E|+{\bf q}_2^{2}/m)} +  \int 
\frac{d^D{\bf k}}{(2\pi )^{D}}\frac{2m\, t_2(2|E|+[2{\bf k}^{2}+2{\bf q}_1^{2}+({\bf k}+{\bf q}_1)^{2}]/4m)\;\phi({\bf 
q}_1,{\bf k})}{4m|E|+{\bf k}^{2}+{\bf q}_1^{2}+{\bf q}_2^{2}+({\bf k}+{\bf q}_1+{\bf q}_2)^{2}} \\ \nonumber
- \frac{1 }{2}\,  \sum_{Q} \frac{(2m)^{2} t_2(|E|+{\bar Q}+{\bf Q}^{2}/4m)\,t_2(|E|-{\bar Q}+{\bf Q}^{2}/4m)\,t_4(|{\bf Q}|,{\bar Q})}{(2m(|E|-{\bar Q})+{\bf q}_1^{2}+({\bf Q}+{\bf q}_1)^{2})(2m(|E|+{\bar Q})+{\bf 
q}_2^{2}+({\bf Q}+{\bf q}_2)^{2})}\; + \; ({\bf q}_1 \leftrightarrow {\bf q}_2)
\label{}
\end{eqnarray}
where the dimer propagator $t_2(x)$ depends on the space dimension $D$. For $D=3$ it is given from Eq.(\ref{2Vertex}) by $t_2(x)\equiv -4\pi /[\,m^{3/2}(\sqrt{x}-\sqrt{E_b)}\,]$, while for $D=2$ according to Eq.(\ref{T2_2D}) we have $t_2(x)\equiv -4\pi /[\,m \ln(x/E_b)]$. 
In the third term the angular 
integration can be performed analytically, and one is left with double integrals for the last two terms, for the $3D$ as well as for the $2D$ case. It is actually quite convenient, in the last term, to deform the 
${\bar Q}$ contour from $]-\infty,\infty[$ to $]-i\infty,i\infty[$ by rotating it by $\pi /2$. No singularity is met in 
this deformation, and one is left to deal only with real quantities.

The above equation has to be supplemented by a corresponding equation for $t_4(|{\bf q}|,{\bar q})$ 
obtained from the definition Eq.(\ref{PhitoT_4}). The important point is that the additional integrations
can be performed analytically, owing to the various invariances under rotations found in the resulting terms. We just give here as an intermediate step the structure of the resulting equation:
\begin{eqnarray}
t_4(k,iz) = S(k,z) +  \int_{0}^{\infty}\!\! \!dp_1\!\int_{0}^{\infty}\!\! \!dp_2\!\int_{0}^{2\pi } \!\!\!
d\alpha 
\;I(k,z,p_1,p_2,\alpha )\; t_2(2|E|+[3{\bf p}_1^{2}+
3{\bf p}_2^{2}+2{\bf p}_1.{\bf p}_2)^{2}]/4m)\,\,\phi({\bf p}_1,{\bf p}_2) \\ \nonumber
+  \int_{0}^{\infty} \!\!\!dK \!\int_{0}^{\infty} \!\!\!dZ \;J(k,z,K,Z)\;|t_2(|E|+iZ+K^{2}/4m)|^{2}\;  t_4(K,iZ)
\label{}
\end{eqnarray}
where $\alpha $ is the angle between ${\bf p}_1$ and ${\bf p}_2$. Here $S(k,z)$, $I(k,z,p_1,p_2,\alpha )$ and $J(k,z,K,Z)$ are analytically known functions of the variables (except that  
$J$ requires to perform numerically a simple integration to be obtained, see below). In this equation and in 
particular in its last term, we have already gone to the purely imaginary frequency variable for $t_4$. The 
resulting $t_4(x,iz)$ turns out to be real and even with respect to $z$.

To be fully specific let us now give the actual self-contained integral equations which we have solved. We restrict 
ourselves to the $3D$ case and to the \emph{bfbf} case (implying $\alpha =1$), corresponding to the 
dimer scattering problem 
treated \cite{Petrov/Salamon:2004,Petrov/Salamon:2005} 
by Petrov {\it et al}. The only generalization is that we keep $E= -\epsilon |E_b|$, instead of setting $\epsilon =1$ as we
should if we considered only the scattering length problem. For clarity we write the resulting equations
with dimensionless quantities, where $1/a$ has been taken as unit wavevector, and $|E_b|=1/ma^2$ as energy unit. For simplicity we keep basically the same notations for the various variables. We just indicate by a 
bar over the function name that they are expressed in reduced units, with reduced variables
(actually we write $\bar{t}_4(k,z)$ instead of $\bar{t}_4(k,iz)$, and there is a change of sign between $\phi({\bf q}_1,{\bf q}_2)$ and $\bar{\phi}({\bf p}_1,{\bf p}_2)$). Equations for other cases and dimensions are completely similar with only few changes in coefficients, signs (for
the particle statistics), for the expression of $\bar{t}_2(x)$ and for the explicit functions coming from analytical angular integrations.

We obtain:
\begin{multline}\label{eqbarphi}
\bar{\phi}({\bf p}_1,{\bf p}_2) = \frac{1}{(\epsilon +{p}_1^{2})(\epsilon +{p}_2^{2})} + \frac{1}{\pi } \int_{0}^{\infty} \!\!k^2\,dk  \int_{0}^{\pi}\!\! \sin\theta\,d\theta \,\frac{\bar{\phi}({\bf p}_1,{\bf k})\,\bar{t}_2(2\epsilon +(3p_1^{2}+3k^{2}+2kp_1\cos\theta)/4)}{\sqrt{A_{+}A_{-}}} + ({\bf p}_1 \leftrightarrow {\bf p}_2) \\
+ \frac{8}{\pi p_1 p_2} \int_{0}^{\infty}\!\!dz\, \int_{0}^{\infty}\!\!dk\,\,\bar{t}_4(k,z) |\bar{t}_2(\epsilon+k^{2}/4 + iz)|^{2}\,I(B_1,B_2,\alpha )
\end{multline}
with $A_{\pm}=2\epsilon +p^{2}_{1}+p^{2}_{2}+k^2+p_1 p_2 \cos\alpha+kp_1 \cos\theta+kp_2 \cos(\alpha \pm \theta)$, and $\alpha $ is the angle between ${\bf p_1}$ and ${\bf p_2}$, while $\theta$ is the polar angle of ${\bf k}$ with ${\bf p_1}$. We have simply set now $\bar{t}_2(x)= [1-\sqrt{x}]^{-1}$.
Here we have also defined the function:
\begin{gather}\label{}
I(B_1,B_2,\alpha )= \Re \frac{1}{2\sqrt{E}} \ln \frac{B_1 B_{2}^{*}+\cos \alpha +\sqrt{E}}{B_1 B_{2}^{*}+\cos \alpha -\sqrt{E}} \\
E=B^{2}_{1}+B^{*2}_{2}+2B_{1}B^{*}_{2} \cos \alpha  - \sin^2 \alpha \\ B(p,k,z)=\frac{1}{kp}\,[\epsilon +p^{2}+\frac{k^2}{2} - iz] \\
B_i \equiv B(p_i,k,z)
\end{gather}

The corresponding equation for $\bar{t}_4(k,z)$ is:
\begin{multline}\label{eqbart4}
\bar{t}_4(k,z)= - \frac{1}{4\pi kz} \,\ln \frac{1+\cos \gamma + 2 \sqrt{\cos \gamma}\cos(\varphi-\gamma/2)}{1+\cos \gamma + 2 \sqrt{\cos \gamma}\cos(\varphi+\gamma/2)} \\
- \frac{1}{\pi ^3 k^2} \int_{0}^{\infty}\!\!p_1\, dp_1 \int_{0}^{\infty}\!\!p_2\, dp_2  \int_{0}^{\pi}\!\! \sin \alpha \,d\alpha \; \bar{\phi}({\bf p}_1,{\bf p}_2)\, \bar{t}_2(2\epsilon +(3p_1^{2}+3p_2^{2}+2p_1p_2\cos\alpha)/4) I(B_1,B_2,\alpha )\\
-\frac{1}{2\pi ^3 k}  \int_{0}^{\infty}\!\!K\,dK  \int_{0}^{\infty}\!\!dZ \,\bar{t}_4(K,Z)
|\bar{t}_2(\epsilon +K^{2}/4 + iZ)|^{2} {\bar J}(k,z,K,Z)
\end{multline}
with $\varphi=\arctan(k/2)$ and $\gamma=\arctan[4z/(4+k^2)]$ and we have defined the function:
\begin{gather}\label{}
{\bar J}(k,z,K,Z)= \int_{0}^{\infty}\!\!dx \;\frac{1}{\epsilon + x^2 +\frac{k^2+K^2}{4}}
\ln \frac{C(x,k,K,Z)}{C(x,-k,K,Z)} \ln \frac{C(x,K,k,z)}{C(x,-K,k,z)}\\
C(x,k,K,Z)=[\epsilon + (x+\frac{k}{2})^2 +\frac{K^2}{4}]^2 + Z^2
\end{gather}

It is seen on these integral 
equations for our two unknown functions $\bar{t}_4(x,z)$ and $\bar{\phi}({\bf p}_1,{\bf p}_2) $ that they require only at 
most a triple integrals to be performed numerically. In this sense they are not numerically more complicated than the work involved in solving directly for the corresponding Schr\"{o}dinger equation, as 
it has been done \cite{Petrov/Salamon:2004,Petrov/Salamon:2005} 
by Petrov {\it et al}. Indeed these integrals require only a
few appropriate change of variables to take care of singular behaviours occuring on some boundaries.
Otherwise they have been performed with unsophisticated integration routine.

In the case of the scattering length a mere iteration algorithm has been found to lead rapidly to the
solution (provided an appropriate exact algebraic manipulation is made to make the iteration convergent). In this way we have been able to handle $45\times45\times45$ matrices (for the three variables entering $\bar{\phi}({\bf p}_1,{\bf p}_2))$. 
This size is large enough to allow improved precision
by extrapolation to infinite size, although we have not done it in the present case, but rather for the ground state of the $bbbb$ complex discussed below. This leads to the result $a_B=0.60\,a_F$ in full agreement with Petrov \cite{Petrov/Salamon:2004,Petrov/Salamon:2005} 
{\it et al}, within a quite reasonable computing time on (nowadays) unsophisticated computer. We have not tried to improve
on the accuracy of the result, since there is no basic interest. In the case of the bound states, to be
described below, we have proceeded to a straight diagonalization of the matrix equivalent to the
right hand sides of Eq.(\ref{eqbarphi}) and Eq.(\ref{eqbart4}) with the Lapack library algorithm. In the 2D case, it is worth noticing that, because of the logarithmic
dependence of $\bar{t}_2(x)$ on $x$, it is quite an improvement to make the change of variables $K=\epsilon ^{1/2} K'$ and $Z = \epsilon Z'$, and so on, since the more appropriate variable turns out to be $\ln \epsilon $ rather than $\epsilon $ itself.

\section{New results in a 2D case}
We will now apply the diagrammatic approach developed in the
previous sections (see also Appendix~\ref{Append1}) to get new
results for the systems of resonantly interacting particles in a
2D case.

As it was first shown by Danilov~\cite{Danilov:1961} (see also a
paper by Minlos and Fadeev~\cite{Minlos/Fadeev:1961})  in the 3D
case, the problem of three resonantly interacting bosons could not be
solved in the resonance approximation. This statement stems from
the fact that in the case of identical bosons the homogeneous part
of Skorniakov-Ter-Martirosian equation~\eqref{a_3} has a non-zero
solution at any energies. The physical meaning of this
mathematical feature was elucidated by Efimov, who showed that a
two-particle interaction leads to the appearance of an attractive
$1/r^2$ interaction in a three-body system. Since in the
attractive $1/r^2$ potential a particle can fall into the center,
the short range physics is important and one can not replace the
exact pair interaction by its resonance approximation.

On the contrary in the case of the 2D problem the phenomena of the
particle fall into the center is absent and one can utilize the
resonant approximation~\cite{Bruch/Tjon:1979,Fedorov:2004}.
Therefore it is possible to describe three- and four-particle
processes in terms of the two-particle binding energy $E_b =
1/ma^2$ only (below, for simplicity we will assume that all
particles under consideration have the same mass $m$). We will
leave aside the problem of composite particles scattering and will
concentrate on the problem of binding energies of complexes of
three and four particles.

As well as in the case of the 3D problem, the cornerstone in the
diagrammatic technique is the two-particle resonance scattering
vertex $T_2$ (see Fig.\ref{fig:Gamma2}). For two resonantly
interacting particles with total mass $2m$ it reads in 2D:
\begin{equation}
\label{T2_2D}
T_2(P) = -\frac{4\pi}{m}\frac{\alpha}{\ln\left(\{\mathbf{P}^2/4m - E\}/|E_B|\right)},%
\end{equation}
where we introduce a factor $\alpha=\{1,2\}$ in order to take into
account whether two particles are indistinguishable or not. That
is $\alpha = 2$ for the case of a resonance interaction between
identical bosons, while $\alpha = 1$ for the case of a resonance
interaction between fermion and boson, or for the case of two
distinguishable fermions.

\subsection{Three particles in 2D}
We start with a system of three resonantly interacting identical
bosons - $bbb$ - in 2D. An equation for the dimer-boson scattering
vertex $T_3$ which describes interaction of three bosons has the
same diagrammatic form as the one shown on the
Fig.\ref{fig:Gamma3}, however there are small changes in the rules 
for its analytical evaluation. The resulting equation reads:
\begin{equation}\label{3Vertex_2D}
    T_3(p_1,p_2;P) = G(P-p_1-p_2)+\sum_{q}G(P-p_1-q)G(q)\,
    T_2(P-q)\;T_3(q, p_2; P),
\end{equation}
where we have now $\sum\limits_q\equiv i \int d^2qd\Omega/(2\pi)^3$, $P = \{\mathbf{0},
E\}$, and one should put $\alpha = 2$ for the two-particle vertex
$T_2$ in~ Eq.\eqref{T2_2D}.  The opposite signs in
Eq.~\eqref{3Vertex} for fermions and Eq.~\eqref{3Vertex_2D} for
bosons are due to the permutational properties of the involved particles
: an exchange of fermions (see Fig.\ref{fig:Gamma3_0})
results in a minus sign, while an analogous exchange of bosons
brings no extra minus. Finally, as we mentionned above, we note that three-particle s-wave
(s-wave channel of a boson-dimer scattering) binding energies
$E_3$ correspond to the poles of $T_3(0,0; \{{\bf 0},-|E_3|\})$ and,
consequently, at energies $E=E_3$ the homogeneous part of
Eq.~\eqref{3Vertex_2D} has a non-zero solution.
Solving Eq.~\eqref{3Vertex_2D} we find that a complex of three
identical bosons has two s-wave bound states $E_3 = -16.5\,E_b$
and $E_3 = - 1.27\,E_b$ in accordance with the previous results
of Bruch and Tjon \cite{Bruch/Tjon:1979,Fedorov:2004}.

Let us now consider a complex - $fbb$ - consisting of one fermion
and two bosons. As noted above we take bosons and fermions with
equal masses $m_{b} = m_{f} = m$. We assume that a fermion-boson
interaction $U_{fb}$, characterized by the length $r_{fb}$, yields
a resonant two-body bound state with an energy $E = -E_b$. In
the same time a boson-boson interaction $U_{bb}$, characterized by
the interaction length $r_{bb}$, does not yield a resonance. Hence
if we are interested in the low-energy physics the only relevant
interaction is $U_{fb}$ and we can ignore the boson-boson
interaction $U_{bb}$, the latter would give small corrections of
the order $|E_B|mr_{bb}^2 \ll 1$ at low energies. In order to
determine three-particle bound states one has to find poles in the
dimer-boson scattering vertex $T_3$. Since we neglect the
boson-boson interaction $U_{bb}$ the vertex $T_3$ is described by
the same diagrammatic equation of Fig.~\ref{fig:Gamma3} as for the
problems of three bosons. The analytical form of this equation
also coincides with Eq.~\eqref{3Vertex_2D} with the minor difference
that the resonance scattering vertex $T_2$ now corresponds to the
interaction between a boson and a fermion, and therefore we
should put $\alpha = 1$ in Eq.~\eqref{T2_2D} for $T_2$.
Solving the equation for $T_3$ we find that $fbb$ complex has only
one s-wave bound state with the energy $E_3 = -2.39\,E_b$.
Note that a complex  - $bff$ - consisting of a boson and two
spinless identical fermions with resonance interaction $U_{fb}$ does not
have any three-particle bound states.

\subsection{Four particles in 2D}
After solving the above three-particle problems we may proceed to
the complexes consisting of four particles. At first we will
consider four identical resonantly interacting bosons
$bbbb$~\cite{Platter:2004}. Any two bosons would form a stable
dimer with binding energy $E= -E_b$. We are going to find a
four-particle binding energy as an energy of an s-wave bound state
of two dimers. Generally speaking a bound state could emerge in
channels with larger orbital moments, however this question will
be a subject of further investigations. Just as in the preceding subsection, 
in order to find a binding energy
we should examine the analytical structure of the dimer-dimer
scattering vertex $T_4$ and find its poles. The set of equations
for~$T_4$ has the same diagrammatic structure as those shown on
Fig.~\ref{fig:PhitoT_4} and Fig.~\ref{fig:Phi}. The analytical
expression for the first equation reads:
\begin{equation}
    \label{T_4_bbbb}
    T_4(p_1,p_2;P)=\frac{1}{\alpha}\sum_{k}
    G(P+p_1-k)G(k)\Phi(P+p_1-k,k;p_2, P),
\end{equation}
and the equation for the vertex $\Phi$ is:
\begin{multline}\label{Phi_bbbb}
    \Phi(q_1,q_2;p_2,P)=G(P-q_1+p_2)G(P-q_2-p_2)
    +\sum_{k}G(k)G(2P-q_1-q_2-k)T_2(2P-q_1-k)\Phi(q_1,k;p_2, P)\\
    +\frac{1}{2\alpha}\sum_{Q, k}G(Q-q_1)G(2P-Q-q_2) T_2(2P-Q)T_2(Q)G(k)G(Q-k)\Phi(k,Q-k;p_2, P)
    +(q_1\leftrightarrow q_2).\\
\end{multline}
where $T_2$ should be taken from Eq.\eqref{T2_2D} and one should
put $\alpha=2$ for the case of identical resonantly interacting
bosons. When we look for the poles of $T_4$ as a
function of the variable $E$, with $P = \{{\bf 0}, E\}$, we have naturally
to consider only the homogeneous part of this equation.
We have found 2 bound states
for the $bbbb$ complex. The values of the total binding energy $|E_{4}|=2|E|$
are given in Table~1 below. Certainly for the validity
of our approximation we should have $|E_4|\ll 1/mr_0^2$. For the
case of four bosons $bbbb$ it means that $197\,E_b \ll1/mr_0^2$ and
hence $a/r_0 \gg \sqrt{197}$. This case can still be considered as quite realistic for
the Feshbach resonance situation.

The case of a four-particle complex -
$bf_{\uparrow}bf_{\downarrow}$ - consisting of resonantly
interacting bosons and fermions is still described by the same
equations~(\ref{T_4_bbbb},\ref{Phi_bbbb}) but with parameter
$\alpha = 1$. In this case we found 2 bound states and they are
also listed in Table~1.

In order to obtain bound states of the $fbbb$ complex one has to
find energies $P = \{\mathbf{0}, E\}$ corresponding to nontrivial solutions
of the following homogeneous equation
\begin{equation}
    \Phi(q_1,q_2;p_2,P)=\sum_{k}G(k)G(2P-q_1-q_2-k)\;T_2(2P-q_1-k)\,\Phi(q_1,k;p_2,P)+(q_1\leftrightarrow q_2).
    \label{eqfbbb}
\end{equation}
This equation corresponds to the diagram of Fig.~\ref{fig:Phi}b.
We have found a single bound state for this $fbbb$ complex.
Finally we summarize the results concerning binding energies of
three and four resonantly interacting particles in 2D in Table~1.
\\
\begin{center}
\begin{tabular}{|c|c|c|c|c|}
\multicolumn{5}{c}{\textbf {Table 1. Bound states of resonantly
interacting particles in 2D}}\\
\hline
  \,System\, & \begin{tabular}{c}
                \rule{0pt}{10pt}Relative$^{1)}$\\interaction\,
                \end{tabular}   & \begin{tabular}{c}
                                  \rule{0pt}{10pt}Number of\\
                                  bound states
                                  \end{tabular} &\,Energy (in
                                  $|E_B|$)$^{2)}$
                                  \,&$\alpha^{3)}$\\

\hline $bbb$  & $U_{bb}$ & 2 & 1.27 , 16.5 & 2 \\
\hline $fbb$  & $U_{fb}$ & 1 & 2.39        & 1 \\
\hline $fbbb$ & $U_{fb}$ & 1 & 4.1        & 1 \\
\hline $bf_{\uparrow}bf_{\downarrow}$
              & $U_{fb}$ & 2 & 2.8, 10.6 & 1 \\
\hline $bbbb$ & $U_{bb}$ & 2 & 22. , 197. & 2 \\

\hline
  \multicolumn{5}{l}{$^1$\rule{0pt}{12pt}\footnotesize Interaction that yields resonance scattering.
  All other interactions are negligible.} \\
  \multicolumn{5}{l}{$^2$\rule{0pt}{12pt}\footnotesize $m=m_b=m_f$.} \\
    \multicolumn{5}{l}{$^3$\footnotesize The indistinguishability parameter in Eq.~\eqref{T2_2D}.}
\end{tabular}
\end{center}

\begin{figure*}
   \scalebox{1.2}{\includegraphics[width=20pc]{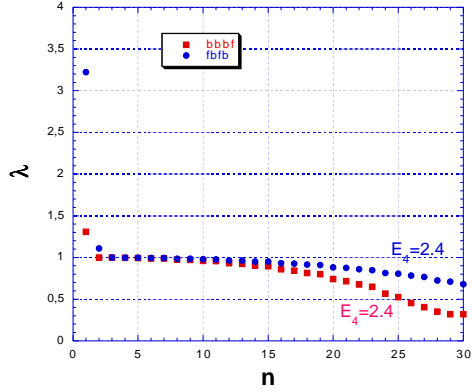}}
   \caption{\label{figbbbf+fbfb}Eigenvalues $\lambda$ found for $|E_{4}|=2.4$, both 
   for the $bbbf$ case
and the $bf_{\uparrow}bf_{\downarrow}$ case. When one of the eigenvalues
is equal to 1, $E_{4}$ is the energy of a possible eigenstate of the complex. The number $n$
appearing on the $x$ axis is just here to number the first few eigenvalues which are displayed by decreasing order.}
\end{figure*}
\begin{figure*}
   \scalebox{1.2}{\includegraphics[width=20pc]{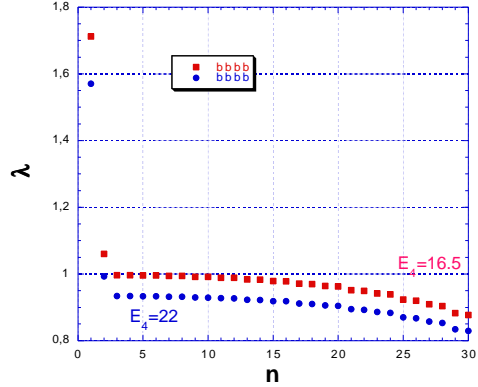}}
   \caption{\label{figbbbb}Eigenvalues $\lambda$ found for $|E_{4}|=16.5$ and
   $|E_{4}|=22.$, for the $bbbb$ case. When one of the eigenvalues
is equal to 1, $E_{4}$ is the energy of a possible eigenstate of the complex. The number $n$
appearing on the $x$ axis is just here to number the first few eigenvalues which are displayed by decreasing order.}
\end{figure*}

For the $bbbb$ complex we find the beginning of a continuum of
states at $|E_{4}|/E_b=16.5$, as it should be since this is, within our
numerical precision, the binding
energy of $bbb$. Similarly we find the beginning of a continuum
at $|E_{4}|/E_b=2.4$ for the $fbbb$ and the $bf_{\uparrow}bf_{\downarrow}$
complex, in agreement with the binding energy of $fbb$. We display our 
corresponding results in Fig.~\ref{figbbbf+fbfb} and Fig.~\ref{figbbbb}.
In all our calculations we find numerically, as a function of $|E_{4}|$, the eigenvalues $\lambda$ corresponding
to the matrix on the right-hand side of our equations, for example Eq.~\eqref{eqfbbb}. 
When one of these eigenvalues is equal to 1, this means that the corresponding
$E_{4}$ is the energy of an eigenstate of our complex. In Fig. \ref{figbbbf+fbfb}, we
display the first highest eigenvalues for $|E_{4}|=2.4$, both for the 
$bf_{\uparrow}bf_{\downarrow}$ case and the $bbbf$ case. One sees clearly
that a fair number of eigenvalues are essentially equal to 1. One could tune
them exactly to 1 by changing very slightly $|E_{4}|$. Hence this corresponds to
the beginning of the continuum. By contrast one sees also clearly two isolated
eigenvalues larger than 1, for the $bf_{\uparrow}bf_{\downarrow}$ case, and
one eigenvalue larger than 1 for the $bbbf$ case. One can bring them to
$\lambda=1$ by increasing $|E_{4}|$, and therefore they correspond to the
bound states that we have found. Similarly we display in Fig. \ref{figbbbb}
the eigenvalues for the $bbbb$ case, for the value $|E_{4}|=16.5$ corresponding
essentially to the threshold for the continuum. Here again one sees many
eigenvalues quite close to 1. On the same figure we also show the results
of the same calculations for $|E_{4}|=22.$ in order to display the way in which
this whole spectrum evolves with $|E_{4}|$. In particular one sees clearly the
two isolated eigenvalues, corresponding to the two bound states found
in this case. In particular since one of them is equal to 1, this means that
the binding energy of one of the bound states is equal to $22\,E_b$, within
our numerical precision.
%
%

Note finally that all our calculations correspond to the case of particles
with equal masses $m_f= m_b = m$, although they can be quite easily
generalized to the case of different masses.

\section{Conclusions and discussion}
For the problem of resonantly interacting fermions in 3D we have
developed an exact diagrammatic  approach that allows to find the
dimer-dimer scattering length $a_B = 0.60 a_F$ in exact agreement
with known results. 
This exact diagrammatic solution of the dimer-dimer scattering length problem  in 
3D opens new horizons for the extension of the self-consistent mean-field 
schemes of Leggett and Nozi\`eres-Schmitt-Rink to the inclusion of the quite essential 
three and four-particle physics in the two-particle variational wave-functions of the 
BCS-type. This in turn will help us to get diagrammatically exact results for $T_c$, 
pseudogap and sound velocity in the dilute BEC-limit and to develop a more 
sophisticated interpolation scheme for these quantities toward the unitarity limit. The work 
on this very exciting project is now in progress.

We have applied the developed approach to get new results in the 2D case.
Namely, we have calculated exactly the binding energies of the following
complexes: three bosons $bbb$, two bosons plus one fermion $bbf$,
three bosons plus one fermion $bbbf$, two bosons plus two fermions
$bf_{\uparrow}bf_{\downarrow}$, and four bosons $bbbb$.

Our investigations enrich the phase-diagram for ultracold
Fermi-Bose gases with resonant interaction. They serve as an
important step for future calculations of the thermodynamical
properties and the spectrum of collective excitations in different
temperature and density regimes, in particular in the superfluid domain. 
Note that in purely bosonic
models in 2D or in the Fermi-Bose mixtures in the case of
prevailing density of bosons $n_B>n_F$ a creation of larger
complexes consisting of 5, 6 and so on particles is also possible.
In fact here we are dealing with the macroscopic phase separation
(with the creation of large droplets). The radius of this droplet
$R_N$ for $N$ bosons in 2D is estimated in~\cite{Hammer:2004} on
the basis of a variational approach. Note that already for $N=5$
the exact calculation of the bound state requires huge
computational capability, but it would be interesting to see
precisely how this would appear with our approach.

\appendix
\section{Dimer-dimer scattering equation. Frequency integration}
\label{Append1}

In this Appendix we will show how one can integrate explicitely over the frequency
dependence in the  dimer-dimer scattering equation~\eqref{Phi} (we consider only
this case, the other ones considered in Section IV would require trivial modifications).
To simplify further computations we slightly change the
notation and introduce a chemical potential $\mu = -E_b/2$ and the single fermion
energy $\xi_{\mathbf{p}}= \mathbf{p}^2/2m - \mu = \mathbf{p}^2/2m + E_b/2$,
with the modified fermion  Green's function $\mathcal{G}(p) =1/(\omega -\xi_{\mathbf{p}})$.
In the expression Eq.(\ref{2Vertex}) for $\TT(Q)$ we have similarly to replace $E$ by
$E-E_b$.

The integral equation~\eqref{Phi} reads more explicitely (with $k = \{\textbf{k},\omega\}$ and
$Q = \{\textbf{Q},\Omega\}$):
\begin{multline}
\label{Phi_exclusion} \TF(q_1,q_2)=-\TG(-q_1)\TG(-q_2)-
    i\int\limits_{-\,\infty}^{\,\infty}\frac{d\omega}{2\pi}\int\frac{d^3\mathbf{k}}{(2\pi)^3}
    \TG(k)\TG(-q_1-q_2-k)\TT(-q_1-k)\TF(q_1,k)    +\\
    +\frac{1}{2}\int\frac{d^4Q}{(2\pi)^4}\frac{d^4k}{(2\pi)^4}\,
    \TG(Q-q_1)\TG(-Q-q_2)
    \TT(-Q)\TT(Q)\TG(k)\TG(Q-k)\TF(k,Q-k)
    +(q_1\leftrightarrow q_2).
\end{multline}
From this equation $\TF(q_1,q_2)= \TF(q_2,q_1)$, as it is obvious physically.
Note also that the third term is already explicitely symmetrical in $q_1\leftrightarrow q_2$.

First we note that, from Eq.(\ref{Phi_exclusion}) itself, $\TF(q_1,q_2)$
is analytical with respect to the frequency variables $\omega _1$ and
$\omega _2$ of the four-vectors $q_1$ and $q_2$ in the lower half-planes
$\Im \,\omega _1 < 0$ and $\Im \,\omega _2 < 0$. This can be seen
by assuming this property self-consistently in the right-hand side, and
checking that the three terms are then indeed analytical, or equivalently 
one can proceed to a perturbative expansion. Then, if we make the "on
the shell" calculation of $\TF(q_1,q_2)$ from Eq.(\ref{Phi_exclusion}),
that is for $\omega_1 = \xi_{\mathbf{q}_1}$ and $\omega_2 = \xi_{\mathbf{q}_2}$,
we see that, for second term in the right-hand side, the only singularity in the
lower complex plane $\Im \,\omega < 0$ is the pole of  $\TG(k)$ at $\omega = 
\xi_{\mathbf{k}}$. Hence the integration contour can be closed in the lower 
half-plane, leading to:
\begin{equation}\label{Term2}
    i\int\limits_{-\,\infty}^{\,\infty}\frac{d\omega}{2\pi}
    \,\TG(k)\TG(-q_1-q_2-k)\TT(-q_1-k)\TF(q_1,k) =
   -\; \frac{\TT\left(-\xi_{\mathbf{q}_1}-\xi_{\mathbf{k}},\mathbf{q}_1+\mathbf{k}\right)}
   {\xi_{\mathbf{q}_1}+ \xi_{\mathbf{q}_2}+
    \xi_{\mathbf{k}}+\xi_{\mathbf{q}_1+\mathbf{q}_2+\mathbf{k}}}\TF(\mathbf{q}_1,\mathbf{k}).
\end{equation}
Here we denote $\TF(\mathbf{q}_1,\mathbf{q}_2) =
\TF(\{\mathbf{q}_1,\xi_{\mathbf{q}_1}\},\{
\mathbf{q}_2,\xi_{\mathbf{q}_2}\})$.

The frequency integration of the third term in Eq.\eqref{Phi_exclusion}
over the frequencies $\Omega$ and $\omega$ is more difficult because
singularities are not essentially located in one half of the complex plane,
as it was the case for the second term. For example $\TF(k,Q-k)$ has
singularities in both half planes, with respect to $\omega $, and similarly
for $\TT(-Q)\TT(Q)$ with respect to $\Omega $. We solve this problem by
splitting the involved functions as the sum of two parts, one analytical in
the upper complex plane, and the other one in the lower complex plane.

First we write $F(\Omega,\textbf{Q},\textbf{q}_1,\textbf{q}_2) \equiv 
\TG(Q-q_1)\TG(-Q-q_2) \TT(-Q)\TT(Q)+(q_1\leftrightarrow q_2)$ 
(we take into account that
we want to calculate $\TF(q_1,q_2)$"on the shell") as:
\begin{eqnarray}
\label{eqF}
F(\Omega,\textbf{Q},\textbf{q}_1,\textbf{q}_2)=
U_+(\Omega,\textbf{Q},\textbf{q}_1,\textbf{q}_2)+
U_-(\Omega,\textbf{Q},\textbf{q}_1,\textbf{q}_2)
\end{eqnarray}
where $U_+$ and $U_-$ are respectively analytical in the upper
and lower complex planes of $\Omega $. This is done by making use of
the Cauchy formula $f(\Omega )=(1/2i\pi )  \int_{C}dz\,f(z)/(z-\Omega )$
for a contour $C$ which encircles the real axis (on which $F$ has no
singularity) and is infinitesimally near of it. This gives:
\begin{eqnarray}
U_+(\Omega,\textbf{Q},\textbf{q}_1,\textbf{q}_2) = \frac{1}{2i\pi }
 \int_{-\infty}^{\infty} dz\, \frac{F(z,\textbf{Q},\textbf{q}_1,\textbf{q}_2)}
 {z-i\epsilon -\Omega }
\label{cauchy}
\end{eqnarray}
with $\epsilon =0_+$. Making use of $F(-\Omega)=F(\Omega)$,
we find $U_-(\Omega,\textbf{Q},\textbf{q}_1,\textbf{q}_2)=
U_+(-\Omega,\textbf{Q},\textbf{q}_1,\textbf{q}_2)$.

On the other hand the last part of the third term $\bar{T}_4(Q') \equiv
 \int d^4k'\,\TG(k')\TG(Q'-k')\TF(k',Q'-k')= \int d^4k'\,\TG(Q'/2+k')\TG(Q'/2-k')
 \TF(Q'/2+k',Q'/2-k')$ satisfies $\bar{T}_4(-Q')=\bar{T}_4(Q')$. This can
 be seen by substituting Eq.\eqref{Phi_exclusion} for $\TF(Q'/2+k',Q'/2-k')$
 in this last expression for $\bar{T}_4(Q')$. For the first term contribution,
 the result is trivial. For the second term, one has to make the shift $
 k \rightarrow k-Q'/2$, and then $k \leftrightarrow k'$. In the third term
 one has to make the shift $k' \rightarrow k'+Q/2$ and then $k' \rightarrow -k'$.
 Then, when we make the change $Q \rightarrow -Q$ in the third term of
 Eq.\eqref{Phi_exclusion} and use $\bar{T}_4(-Q)=\bar{T}_4(Q)$, we see that
 the $U_-$ contribution is exactly identical to the $U_+$ contribution and we are
 left with a single contribution from $U_-$ to evaluate.
 
 In  order to perform the $\omega $ integration in $\bar{T}_4(Q)=
\int d^4k\,\TG(Q/2+k)\TG(Q/2-k) \TF(Q/2+k,Q/2-k)$, we split:
\begin{eqnarray}
\label{eqTF}
\TF(Q/2+k,Q/2-k)=\TF_+(Q/2+k,Q/2-k)+\TF_-(Q/2+k,Q/2-k)
\end{eqnarray}
into the sum of two functions, with $\TF_+$ analytical
in the upper complex plane with respect to $\omega $, and $\TF_-$
analytical in the lower complex plane. That this can be
done is immediately seen from Eq.\eqref{Phi_exclusion} itself. For
the first term we just have to write the product of Green's functions as
$\TG(k-Q/2)\TG(-k-Q/2)=-(\TG(k-Q/2)+\TG(-k-Q/2))/(\Omega+ 
\xi_{\mathbf{k+Q}/2+\xi(\mathbf{k-Q}/2}-i\epsilon )$, which has explicitely the
required property. In the third term we can handle the product of the
first two Green's functions in the same way. Finally, in the second
term, after performing the $\omega $ integration as indicated above
(but without taking the "on the shell" values for the frequencies),
one sees that the result for the term written explicitely above in
Eq.\eqref{Phi_exclusion} is analytical in the lower complex plane
with respect to $\omega $. The corresponding term obtained by
$(q_1\leftrightarrow q_2)$ is analytic in the upper complex plane.
In each case one checks that the functions analytical in the upper
and lower complex plane are related by $k  \leftrightarrow -k$, so
that $\TF_-(Q/2+k,Q/2-k)=\TF_+(Q/2-k,Q/2+k)$. Hence by the
change of variable $k  \leftrightarrow -k$, the contributions of
$\TF_+$ and $\TF_-$ are equal. 

Then we have arrived, for the calculation of $\bar{T}_4(Q)$, to a situation
which is similar to the one we met for three particles. Since  $\TF_+(Q/2-k,Q/2+k)$
and $\TG(Q/2-k)$ are analytical in the lower complex plane, we can close the
integration contour at infinity in this lower half plane and the only contribution
comes from the pole of $\TG(Q/2+k)$. This leads to:
\begin{eqnarray}
\bar{T}_4(Q)=-2i \int \frac{d \mathbf{k}}{(2\pi )^3} \frac{\mathcal{F}(\Omega,
\mathbf{k},\mathbf{Q})}{\Omega -\xi_{\mathbf{k+Q}/2}
-\xi_{\mathbf{k-Q}/2}+i\epsilon }
\label{eqT4}
\end{eqnarray}
where $\mathcal{F}(\Omega,\mathbf{k},\mathbf{Q})$ is $\TF_+(Q/2-k,Q/2+k)$ 
evaluated for $\omega = \xi_{\mathbf{k+Q}/2} - \Omega/2$. An important property,
which can be checked on each term contributing to $\TF_+(Q/2-k,Q/2+k)$ is that
$\mathcal{F}(\Omega,\mathbf{k},\mathbf{Q})$ is analytical in the lower complex 
plane with respect to $\Omega$. Hence the integration of $U_-(\Omega,\textbf{Q},\textbf{q}_1,\textbf{q}_2) \bar{T}_4(Q)$ over $\Omega $
can also be performed by closing the contour in the lower half plane, since
the only singularity in this half plane is the pole due to the denominator in 
Eq.(\ref{eqT4}). The contribution of this pole leads to the evaluation of 
$\mathcal{F}(\Omega,\mathbf{k},\mathbf{Q})$ for $\Omega=\xi_{\mathbf{k+Q}/2}
+\xi_{\mathbf{k-Q}/2}$. Taken with the above definition of $\mathcal{F}$ this means 
that we have calculated $\TF_+(Q/2-k,Q/2+k)$ for $\Omega/2 - \omega = 
\xi_{\mathbf{k-Q}/2}$ and $\Omega/2 + \omega = \xi_{\mathbf{k+Q}/2}$, which
is just an evaluation "on the shell". Because of the simple relation between
$\TF_+$ and $\TF_-$ the result can be expressed in terms of
$\TF(\mathbf{k+Q}/2,\mathbf{-k+Q}/2)$ itself.

Gathering all the above results we end up with the following complete equation
for $\TF(\mathbf{q}_1,\mathbf{q}_2)$:
\begin{multline}\label{Phi_final}
\mathit{\Phi}(\mathbf{q_1}, \mathbf{q_2}) = - \frac{1}{4 \xi_{\mathbf{q}_1} \xi_{\mathbf{q}_2}}
+ \int \frac{d^3\mathbf{k}}{(2\pi)^3}\,
\frac{\TT\left(-\xi_{\mathbf{q}_1}-\xi_{\mathbf{k}},\mathbf{q}_1+\mathbf{k}\right)}
{\xi_{\mathbf{q}_1}+\xi_{\mathbf{q}_2}+
\xi_{\mathbf{k}}+\xi_{\mathbf{q}_1+\mathbf{q}_2+\mathbf{k}}}\TF(\mathbf{q}_1,\mathbf{k})\\
-\int\frac{d^3\mathbf{Q}}{(2\pi)^3}\frac{d^3\mathbf{k}}{(2\pi)^3}
U(\xi_{\mathbf{k+Q}/2}+\xi_{\mathbf{k-Q}/2},\textbf{Q},\textbf{q}_1,\textbf{q}_2)
\mathit{\Phi}(\mathbf{k+Q}/2, \mathbf{-k+Q}/2)
+(q_1\leftrightarrow q_2).
\end{multline}
In this equation we have modified the integration contour in the definition of $U_-$ 
to have it running on the imaginary axis rather than on the real axis, and we have
used the symmetry property of $F(z,\textbf{Q},\textbf{q}_1,\textbf{q}_2)$ with respect
to $z$, together with symmetry properties of $\TF(\mathbf{q}_1,\mathbf{q}_2)$,
to rewrite the result in terms of the real function:
\begin{eqnarray}
U(\Omega,\textbf{Q},\textbf{q}_1,\textbf{q}_2) = \frac{\Omega }{\pi }
 \int_{0}^{\infty} dy\, \frac{F(iy,\textbf{Q},\textbf{q}_1,\textbf{q}_2)}
 {y^2+\Omega^2 }
\label{eqU}
\end{eqnarray}
which shows that $\mathit{\Phi}(\mathbf{q_1}, \mathbf{q_2})$ itself is real.

We have made practical numerical use of Eq. (\ref{Phi_final}) to find for example the
ground state energy. Although this turned out to be quite feasable, this equation
appears finally less convenient than what we have described in section \ref{pract}.
This was expected since the solution implies quadruple integrals, instead of the 
triple integrals we had only to deal with in section \ref{pract}.



\section{Modified dimer-dimer scattering equation}
\label{Append2}

This appendix is devoted to an alternative description of the
dimer-dimer scattering process. The purpose is to obtain a
direct integral equation for $T_4(p_1,p_2;P)$, in a way
convenient for numerical calculations. Below we derive such a set
of equations, that were used for practical computations as indicated
in section \ref{pract}.

\begin{figure*}
    \includegraphics[width=30pc]{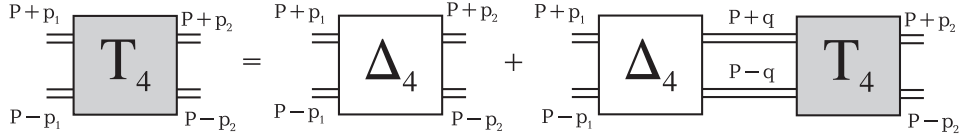}
    \caption{\label{fig:T4} The diagrammatic representation of the equation
    for the full dimer-dimer scattering vertex $T_4(p_1,p_2;P)$.}
\end{figure*}

\begin{figure*}
    \includegraphics[width=15pc]{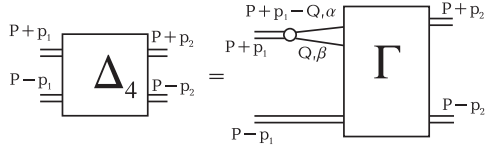}
    \caption{\label{fig:Delta4} The diagrammatic representation of the sum of all
    irreducible diagrams $\Delta_4(p_1,p_2;P)$.}
\end{figure*}

The first step is to construct for
two dimers a "bare" interaction potential, or vertex, $\Delta_4$, 
which is the sum of all irreducible
diagrams, and then to build ladder diagrams from this 
vertex, in order to obtain an integral equation (see Fig.
\ref{fig:T4}). These irreducible diagrams are those ones which
cannot be divided by a vertical line into two parts connected by
two dimer lines. As it was pointed above the vertex $\Delta_4$ is
given by the series shown on Fig.\ref{fig:Gamma4_0}{\it e}, since
the diagrams on Fig.\ref{fig:Gamma4_0}{\it f} are by contrast reducible. Again we
can eliminate $T_3$  from our considerations and express $\Delta_4$
only in terms of $T_2$. For this purpose we have to introduce a
special vertex with two fermionic and one dimer incoming lines and
two dimer outgoing lines $\Gamma_{\alpha\beta}(q_1,q_2;p_2, P)$
(see Fig. \ref{fig:Delta4}). This vertex
$\Gamma_{\alpha\beta}(q_1,q_2;p_2, P)$ corresponds to the vertex
$\Delta_4$ with one incoming dimer line being removed, in much the
same way as $\Phi(q_1,q_2;p_2, P)$ and $ T_4(p_1,p_2;P)$ are
related in Eq. (\ref{PhitoT_4}). The difference is that 
$\Gamma_{\alpha\beta}(q_1,q_2;p_2, P)$ is irreducible with respect
to two dimer lines while $\Phi(q_1,q_2;p_2, P)$ is not, just in the
same way as $T_4(p_1,p_2;P)$ and $\Delta_4(p_1,p_2;P)$ are related.
The corresponding equation relating
$\Gamma_{\alpha\beta}(q_1,q_2;p_2, P)$ and $ \Delta_4(p_1,p_2;P)$ is:
\begin{equation}\label{Delta}
    \Delta_4(p_1,p_2;P)=\frac{1}{2}\sum_{Q;\, \alpha, \beta}\chi(\alpha, \beta)
    G(P+p_1-Q)G(Q)\Gamma_{\alpha\beta}(P+p_1-Q,Q;p_2, P).
\end{equation}

\begin{figure*}
    \includegraphics[width=30pc]{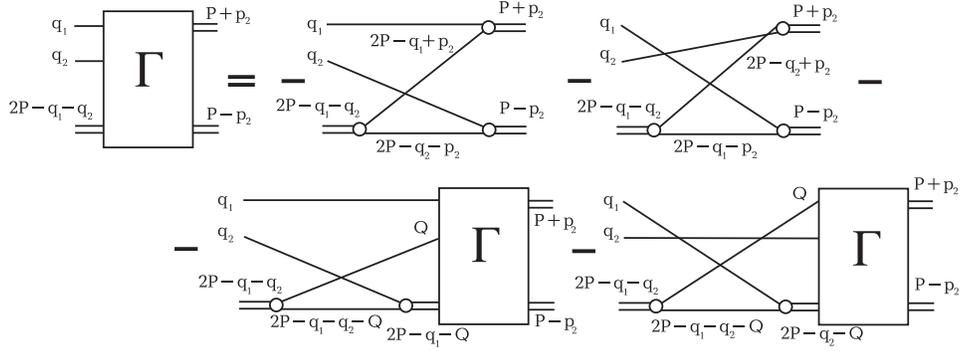}
    \caption{\label{fig:Gamma4} The graphic representation of the equation
    on the full vertex $\Gamma(q_1,q_2;p_2, P)$.}
\end{figure*}

One can readily verify that the diagrammatic expansion for $\Gamma$
shown on Fig.~\ref{fig:Gamma4} yields the same series as the
one shown on Fig.~\ref{fig:Gamma4_0}{\it e} for the vertex
$\Delta_4$. The spin part of $\Gamma_{\alpha, \beta}$ has again the simple
form $\Gamma_{\alpha, \beta}(q_1,q_2;P,p_2) = \chi(\alpha,
\beta)\Gamma(q_1,q_2;p_2,P)$ and the function $\Gamma(q_1,q_2;p_2,
P)$ obeys the following equation:
\begin{multline}\label{Gamma}
    \Gamma(q_1,q_2;p_2, P)=-G(P-q_1+p_2)G(P-q_2-p_2)-G(P-q_2+p_2)G(P-q_1-p_2)-\\
    -\sum_{Q}G(Q)G(2P-q_1-q_2-Q)\left[T_2(2P-q_1-Q)\Gamma(q_1,Q;p_2, P)+
    T_2(2P-q_2-Q)\Gamma(Q,q_2;p_2, P)\right].
\end{multline}
The sign minus in \eqref{Gamma} is a consequence of the
anticommutativity of Fermi operators. It is clear that
Eqs.~\eqref{Delta} and \eqref{Gamma} can be analytically
integrated over the variable $\Omega$. Thus the $s$-wave component of
the vertex $\Gamma(q_1,q_2;p_2, P)$ is a function of the absolute
values of vectors $|\mathbf{q}_1|$ and $|\mathbf{q}_2|$, the angle
between them, the absolute value of vector $|\mathbf{p}_2|$, and
the frequency $\omega_2$. The $s$-wave component of the sum of all
irreducible diagrams $\Delta_4(p_1,p_2;P)$ is a function of the
absolute values of the vectors $|\mathbf{p}_1|$ and
$|\mathbf{p}_2|$ and the frequencies $\omega_1$ and $\omega_2$.

The fully symmetrized vertex $T_4(p_1,p_2;P)$ of two-dimer scattering
can be found from the solution of the following equation (see
Fig. \ref{fig:T4}):
\begin{equation}
    \label{T_4}
    T_4(p_1,p_2;P)=\Delta_4(p_1,p_2;P)+\frac{1}{2}\sum\limits_q
    \Delta_4(p_1,q;P)T_2(P+q)T_2(P-q)T_4(q,p_2;P),
\end{equation}
where $\Delta_4(p_1,p_2;P)$ is the sum of all irreducible diagrams,
$P\pm p_{1,2}=\{-E_b\pm \omega_{1,2},\pm\mathbf{p}_{1,2}\}$ are
4-vectors of incoming (1) and outgoing (2) dimers in the
center-of-mass system.

Let us finally note that, equivalently to our above derivation,
Eq. (\ref{Delta}-\ref{T_4}) can be also related
to Eq. (\ref{PhitoT_4}) and Eq. (\ref{Phi}) algebraically by simple
formal operator manipulations.

\begin{acknowledgments}
This work was  supported by  Russian Foundation for Basic Research
(Grant No.  04-02-16050), CRDF (Grant No. RP2-2355-MO-02) and the
grant of Russian Ministry of Science and Education. We are
grateful to  A.F. Andreev, I.A. Fomin, P. Fulde, Yu. Kagan, L.V.
Keldysh, Yu. Lozovik, S.V. Maleev, B.E. Meierovich, A.Ya. Parshin, P. Pieri
T.M. Rice, V.N. Ryzhov, G.V. Shlyapnikov, G.C. Strinati, V.B.
Timofeev, D. Vollhardt and P. W\"{o}lfle for fruitful discussions.
M.Yu.K is grateful to the University Pierre and Marie Curie for
the hospitality on the first stage of this work.

Laboratoire de Physique Statistique is " Laboratoire associ\'e au Centre National
de la Recherche Scientifique et aux Universit\'es Paris 6 et Paris 7 ".
\end{acknowledgments}

\bibliography{BoseFermi}
\end{document}